\documentstyle[11pt,paspconf,epsf]{article}

\begin{document}

\title{Galaxy Transformation by Merging}

\author{Joshua E. Barnes}

\affil{Institute for Astronomy, University of Hawai`i,
       2680 Woodlawn Drive, Honolulu, Hawai`i, 96822, USA}

\begin{abstract}
Theoretical considerations and observational data support the idea
that mergers were more frequent in the past.  At high redshifts,
violent interactions and mergers may be implicated in the origin of
Lyman-break galaxies, sub-mm starbursts, and active galactic nuclei.
Most stars in cluster ellipticals probably formed at redshifts $z >
2$, as did most of the halo and globular clusters of the Milky Way;
these events may all be connected with mergers.  But {\it what\/} kind
of galaxies merged at high redshifts, and {\it how\/} are these early
events connected to present-epoch mergers?  I will approach these
questions by describing ideas for the formation of the Milky Way,
elliptical galaxies, and populations of globular clusters.
\end{abstract}

\section{Introduction}

Galaxy merging, once controversial, has become all too respectable.  A
recent {\it New York Times\/} article claims ``The importance of
mergers in the evolution of galaxies was one of the insights gained
from the northern [Hubble Deep Field] study'' (Wilford 1998).  But
this insight arrived long before {\it HST\/}'s spectacular
observations.  Theoretical arguments for violent interactions and
mergers between galaxies include:

\begin{itemize}

\item
Hierarchical clustering, in which small objects are progressively
incorporated into larger structures (Layzer 1954), is common to many
accounts of galaxy formation.  In the ``core-halo'' picture (White \&
Rees 1978), clustering of dark matter creates galaxy halos which
subsequently accumulate cores of baryons, forming visible galaxies.

\item
Tidal encounters generate short-lived features; a population of binary
galaxies with highly eccentric orbits is required to explain the
peculiar galaxies observed today (Toomre \& Toomre 1972).  If these
binaries have a flat distribution of binding energies, their merger
rate has declined with time as $t^{-5/3}$, and the $10$ or so merging
galaxies in the NGC catalog are but the most recent additions to a
population of about $750$ remnants (Toomre 1977).

\item
The CDM model (Blumenthal et al.~1984) provides a concrete example of
galaxy formation in which merging of dark halos is easily calculated
and clearly important (Lacey \& Cole 1993).

\end{itemize}

Observations, though until recently limited to redshifts below those
probed in the HDF, also indicate rapid merging at high redshift:

\begin{itemize}

\item
Various counting strategies, mostly limited to $z \la 1$, show the
pair density growing like $(1 + z)^m$, where $m \simeq 3 \pm 1$ (Zepf
\& Koo 1989, Abraham 1999).

\item
Peculiar morphology becomes more common with increasing redshift (van
den Bergh et al.~1996).  For example, the fraction of irregular
galaxies in the CFRS survey increases from about $10$\% at $z \sim
0.4$ to a third at $z \sim 0.8$ (Brinchmann et al.~1998).

\end{itemize}

Thus, both theory and observation support the notion that there was
``a great deal of merging of sizable bits and pieces (including quite
a few lesser galaxies) early in the career of every major galaxy''
(Toomre 1977).  But the {\it nature\/} of these early mergers is not
so clear; were the objects involved dominated by dark matter, by gas,
or by stars?  And can we learn anything about early mergers by
studying present-epoch examples?

\section{Signposts of High-Redshift Merging}

Merging is hard to prove at redshifts $z \ga 1.5$; cosmological
dimming renders tidal tails nearly invisible, while bandshifting
effects complicate interpretation of the observations (Hibbard \&
Vacca 1997).  But circumstantial evidence implicates merging in
various high-$z$ objects.

\subsection{Starburst Galaxies}

The most extensively studied high-redshift galaxies are the
``Lyman-break'' objects at $z \sim 3$, which have rest-frame UV
luminosities consistent with star formation rates of $\sim 10
{\rm\,M_\odot\,yr^{-1}}$ (Steidel et al.~1996b).  The actual rates
could be several times higher, since much of the UV emitted by young
stars may be absorbed by dust (e.g.~Heckman 1998).  Spectra show gas
outflows with velocities of $\sim 500 {\rm\,km\,sec^{-1}}$ (Pettini et
al.~1998a), atypical of quiescent galaxies but fairly normal for
starburst systems.  Heavily obscured high-$z$ starbursts have been
detected at sub-mm wavelengths (Hughes et al.~1998, Barger et
al.~1998).  These have IR spectral energy distributions similar to
those of ultra-luminous starburst galaxies like Arp~220 and appear to
be forming stars at rates of $\sim 10^2 {\rm\,M_\odot\,yr^{-1}}$.

At low redshifts, luminous starbursts are often triggered by mergers
of gas-rich galaxies (Sanders \& Mirabel 1996).  The gas becomes
highly concentrated; H$_2$ surface densities of $10^3$ to~$10^5
{\rm\,M_\odot\,pc^{-2}}$ are typical of nearby starbursts (Kennicutt
1998a), and similar surface densities are indicated in high-$z$
starbursts (Heckman 1998).  In the potential of an axisymmetric
galaxy, gas becomes ``hung up'' in a disk several kpc in radius
instead of flowing inward.  Violently changing potentials in merging
galaxies enable gas to shed its angular momentum and collapse to as
little as $\sim 1$\% of its initial radius (Barnes \& Hernquist 1996).

Models based on mergers of low-$z$ disk galaxies may not be a good
description of high-redshift starbursts.  First, compared to nearby
starbursts, Lyman-break galaxies have higher peak UV surface
brightnesses (Weedman el al.~1998) and seem to be more extended
(Pettini et al.~1998b); these differences may be due to higher gas
contents, lower dust extinction, or undetected AGN.  Second, disks
collapsing at higher redshifts are naturally more compact (Mo, Mao, \&
White 1998) and thus may {\it form\/} with surface densities
characteristic of starbursts.  Third, bar instabilities in isolated
galaxies can drive rapid gas inflows even without external triggers
(Schwarz 1981).  Nonetheless, many high-$z$ objects have highly
irregular shapes, and deep HDF images reveal faint asymmetric features
which may be due to tidal interactions (van den Bergh et al.~1996,
Steidel et al.~1996a).  Mergers seem to be the ``best bet'' for
high-$z$ starbursts (Somerville, Primack, \& Faber 1998), but
extrapolation from low-$z$ is only the first step toward testing this
conjecture.

\subsection{Radio Galaxies}

At low redshifts, powerful radio sources are often associated with
merger remnants; some $30$\% exhibit tails, fans, shells, or other
signatures of recent collisions (Heckman et al.~1986).  But at
redshifts $z \ga 0.6$ the most striking morphological feature of
powerful radio sources is a near-ubiquitous alignment between the
radio lobes and continuum optical emission (McCarthy et al.~1987,
Chambers, Miley, \& van Breugel 1987).  This ``alignment effect''
seems at odds with the merger morphologies seen at low redshift; one
explanation invokes jet-induced star formation (e.g.~McCarthy et
al.~1987).

Recent observations suggest the alignment effect is compatible with
mergers (Stockton 1999).  Strong polarization is found in several $z
\ga 2$ radio galaxies, implying that the aligned emission is scattered
light from an obscured AGN (e.g.~Tadhunter, Fosbury, \& di Serego
Alighieri 1989); in several cases there is good evidence that dust is
the primary scattering agent (Knopp \& Chambers 1997, Rush et
al.~1997).  HST imaging of the radio galaxy 0406--244 at $z = 2.44$
reveals a double nucleus and what appear to be tidal debris
illuminated by an AGN (Rush et al.~1997).

From a theoretical point of view, merging may be the most efficient
way to form powerful radio sources.  The engines of such galaxies are
probably rapidly spinning black holes (Begelman, Blandford, \& Rees
1984).  Accretion from a disk is an ineffective way to spin up a black
hole, since little of the disk's angular momentum falls into the hole;
on the other hand, two black holes of comparable mass can spiral
together to produce a rapidly-spinning hole (Wilson \& Colbert 1995).

\subsection{Quasars}

Evidence that low-redshift quasars frequently occur in interacting
systems has been accumulating for two decades (Stockton 1999).  Early
claims that quasars have close companions are supported by recent
studies out to redshifts $z \sim 1$ (Disney et al.~1995, Fisher et
al.~1996, Stockton \& Ridgway 1998).  Even more telling are the tidal
tails and other signs of violent interactions in nearby cases
(Stockton \& Mackenty 1983, Stockton \& Ridgway 1991, Bahcall,
Kirhakos, \& Schneider 1995, Boyce et al.~1996, Stockton, Canalizo, \&
Close 1998).

The very nature of these interactions makes their detection difficult
at high redshifts -- there, tidal tails and other signs would be
hidden by cosmological dimming and quasar glare.  Nor do the low-$z$
observations preclude the possibility that high-redshift quasars may
have nothing to do with mergers.  A compelling case that high-$z$ AGN
are sparked by mergers probably awaits a theory for the formation of
supermassive black holes.  All the same, it's probably no coincidence
that the peak in quasar activity at $z \sim 2$ to~$3$ broadly
coincides with other indications of rapid merging reviewed above.

\section{Assembling the Milky Way}

Complementing the data gathered by looking back to high redshift is
information gleaned by ``archeological'' studies of objects at $z \sim
0$.  Mergers of small galaxies probably played an important role in
the formation of the Milky Way's halo (Searle \& Zinn 1978); the
evidence includes (cf. Gilmore, these proceedings):

\begin{enumerate}

\item
A ``second parameter'' -- which may (Sarajedini, Chaboyer, \& Demarque
1997) or may not (Stetson, VandenBerg, \& Bolte 1996) be age -- is
required to account for variations in globular cluster horizontal
branch morphologies.

\item
This second parameter is correlated with the shape of a cluster's
orbit; for example, clusters with retrograde orbits have Oosterhoff
class~I variables (van den Bergh 1993).

\item
Halo stars with $[{\rm Fe}/{\rm H}] \sim -1$ have a large range of
$[\alpha/{\rm Fe}]$ values (Gilmore \& Wyse 1998, Stephens 1999).

\item
The outer halo exhibits retrograde rotation with respect to the rest
of the galaxy (Majewski 1996).

\item
Observations of moving groups in the halo (Eggen 1987, Majewski, Munn,
\& Hawley 1994).

\item
The Magellanic stream (e.g.~Mathewson 1985) and other ``ghostly
streams'' of dwarf galaxies and halo globulars (Lynden-Bell \&
Lynden-Bell 1995).

\item
High-latitude A stars in the halo (Preston, Beers, \& Schectman 1994).

\item
The Sgr~I dwarf galaxy, apparently being torn apart by the Milky Way
(Ibata, Gillmore, \& Irwin 1995).

\end{enumerate}

The variety of stellar populations and abundance patterns (items 1--3)
indicate that different parts of the halo have different enrichment
histories.  Moreover, the strong correlation between Oosterhoff class
and orbital direction (item~2) suggests that at least one object with
a mass of $\sim 10^{10} {\rm\, M_\odot}$ fell in on a retrograde orbit
(van den Bergh 1993), and the rotation of the outer halo (item~4)
likewise implies a fairly massive retrograde component.  Moving groups
(item~5) indicate that the stellar halo formed from kinematically
distinct components and is not yet well-mixed.  Incomplete mixing is
also implied by tentative identifications of ghostly streams (item~6),
though the interpretation of a stream as the remnant of a single
tidally-disrupted galaxy is problematic since {\it multiple\/} dark
halos apparently exist in each stream (Kormendy 1990).  Streams may
instead trace infalling filaments; these may be associated with
high-velocity clouds (Blitz et al.~1999).  The A stars at high
latitudes (item~7) are too young to have been scattered from the
galactic disk and may therefore represent fairly recent acquisitions.
Finally, the Sgr~I dwarf galaxy (item~8) provides a clear example of
halo accretion as an ongoing process.

Two different arguments suggest that the {\it bulk\/} of the halo fell
into place long ago.  First, {\it most\/} halo stars are old.  The
halo as a whole shows a well-defined turn-off at $B - V \sim 0.4$,
corresponding to ages $\ga 10 {\rm\,Gyr}$; only $\sim 10$\% of the
stars appear younger (Unavane, Wyse, \& Gilmore 1996).  To be sure,
this does not rule out recent accretions of objects containing only
old stars, but most dwarf galaxies in the local group contain
intermediate-age stars as well.  Thus, unless the accreted galaxies
were unlike those we observe today, most fell in more than $10
{\rm\,Gyr}$ ago.

Second, galactic disks are dynamically fragile; accretion of satellite
galaxies can easily ruin a stellar disk.  Analytic estimates limit the
mass accreted by the Milky Way to less than $4$\% in the past $5
{\rm\,Gyr}$ (T\'oth \& Ostriker 1992).  N-body experiments show about
half the disk heating that analytic work predicts; dark halos absorb
much of the damage, and disks may tilt as well as thicken (Walker,
Mihos, \& Hernquist 1996, Huang \& Carlberg 1997, Vel\'azquez \& White
1998).  Still, accretion events of any size increase the disk's
vertical dispersion, $\sigma_{\rm z}$.  Signatures of past mergers may
be sought in the $\sigma_{\rm z}$--age relation; most striking is the
jump from $\sigma_{\rm z} \simeq 20$ to~$40 {\rm\,km\,sec^{-1}}$ which
probably marks the transition to the $\sim 10 {\rm\,Gyr}$-old thick
disk (e.g.~Gilmore, Wyse, \& Kuijken 1989, Freeman 1993).

Thus it appears that the Milky Way last suffered a significant merger
at least $10 {\rm\,Gyr}$ ago; relics of this event include the outer
stellar halo and the thick disk.  Presumably, the Milky Way's dark
halo was largely in place at this time, since a major merger would
have disrupted even the thick disk.

\section{Assembling Early-Type Galaxies}

While disk galaxies like the Milky Way have had fairly quiet lives for
the past $10 {\rm\,Gyr}$, some elliptical and S0 galaxies have more
complex histories.  As a group, elliptical galaxies appear to have
been shaped by violent relaxation, and almost any plausible account
for such violence involves the coalescence of multiple lumps -- that
is, some sort of merger.  But {\it when\/} were these mergers, and
{\it what\/} sort of objects were involved?

\subsection{Merger Formation}

After much debate, it's clear that elliptical galaxies include some
objects formed by fairly {\it recent\/} mergers of disk systems.
Support for this position includes:

\begin{itemize}

\item
The merger origin of ultra-luminous IR galaxies (e.g.~Sanders \&
Mirabel 1996), and the suggestion that such events could build the
central regions of elliptical galaxies (Schweizer 1990, Kormendy \&
Sanders 1992).

\item
Studies of young merger remnants like NGC~7252 (Schweizer 1982) and
models of disk galaxy mergers reproducing such objects (Barnes 1988,
Hibbard \& Mihos 1995).

\item
Evidence for recent star formation in ellipticals, as indicated by
H$\beta$ line strengths (Faber et al.~1994).

\item
Correlations between ``fine structures'' in elliptical galaxies and
residuals in the luminosity--color and luminosity--line strength
relations (Schweizer et al.~1990, Schweizer \& Seitzer 1992).

\end{itemize}

These results trace the gradual assimilation of recent merger remnants
into the general population of early-type galaxies.  Estimated merger
ages of nearby field Es and S0s, based on luminosity--color residuals
and a color evolution model for disk-galaxy merger remnants, range
from $\sim 10 {\rm\,Gyr}$ down to $3$--$5 {\rm\,Gyr}$, with a handful
of younger objects (Schweizer \& Seitzer 1992).  Complementary
evidence from deep photometric surveys indicates that only a third to
a half of all field ellipticals were in place and passively evolving
by $z \simeq 1$ (Kauffmann, Charlot, \& White 1996, Barger et
al. 1999).

Such evidence is scant for cluster ellipticals, which seem to be an
older and more homogeneous population.  Galaxy clusters are old in two
distinct respects: first, cluster galaxies collapsed early; second,
dynamical processes run faster in proportion to $\sqrt{\rho}$.  A
study of the fundamental plane out to $z \simeq 0.83$ indicates that
most cluster ellipticals have evolved passively since forming the bulk
of their stars at $z \ga 2$ (van Dokkum et al.~1998).

Counter-rotating or kinematically decoupled ``cores'' are probably the
clearest signs that cluster ellipticals were formed by mergers (Surma
\& Bender 1995, Mehlert et al.~1998).  High-resolution imaging shows
that kinematically distinct nuclear components are {\it disks\/}
(Surma \& Bender 1995, Carollo et al.~1997).  These disks typically
have high metal abundances (Bender \& Surma 1992) and low velocity
dispersions (Rix \& White 1992); such properties indicate that they
formed dissipationally during major mergers (Franx \& Illingworth
1988, Schweizer 1990).  Merger simulations producing counter-rotating
nuclear gas disks back up this hypothesis (Hernquist \& Barnes 1991).

The nature of the mergers which formed cluster ellipticals is unclear;
often invoked are highly dissipative encounters of gaseous fragments
(e.g.~Bernardi et al.~1998, Thomas et al.~1999).  But the existence of
counter-rotating disks in cluster Es indicates that their immediate
ancestors can't have been very numerous or very gassy.  If many small
objects coalesced, the law of averages would make counter-rotation
extremely rare.  And counter-rotation is unlikely to arise in
essentially gaseous mergers since gas flows can't interpenetrate.

Relatively few mergers are expected once a cluster has virialized;
encounters at speeds higher than about twice a galaxy's internal
velocity dispersion don't result in mergers (e.g.~Makino \& Hut 1997).
Observations of kinematically distinct disks in cluster galaxies
support this expectation, since such disks are unlikely to survive
major dissipationless mergers (Schweizer 1998).

In sum, merging apparently began early in proto-cluster environments
but tapered off as the clusters themselves virialized, while in field
environments merging began later and has continued up to the present.
This general picture is broadly supported by semi-analytic treatments
of galaxy formation in CDM (Kauffmann 1996, Baugh, Cole, \& Frenk
1996).  If it is correct then systematic differences between cluster
and field ellipticals are expected.  The observational status of this
prediction is unclear; one study finds field Es are bluer, have lower
${\rm Mg}_2$ indices, and higher surface brightnesses than cluster Es
of the same luminosity (de Carvalho \& Djorgovski 1992), while another
study finds essentially identical ${\rm Mg}_2$--$\sigma_0$ relations
for field and cluster Es (Bernardi et al.~1998).

\subsection{Abundance Ratios}

Relative abundances of $\alpha$-process elements with respect to Fe
are several times higher in elliptical galaxies than they are in the
disk of the Milky Way (Worthey, Faber, \& Gonzalez 1992, Davies,
Sadler, \& Peletier 1993).  This may constrain the enrichment history;
high $[{\alpha}/{\rm Fe}]$ ratios favor enrichment by SN~II on a short
timescale, while solar ratios ($[{\alpha}/{\rm Fe}] = 0$) favor
enrichment by both SN~II and SN~Ia on a timescale much longer than $1
{\rm\,Gyr}$.  The high $[{\alpha}/{\rm Fe}]$ ratios seen in elliptical
galaxies indicate that SN~Ia played little role in their chemical
evolution; on the face of it, they also imply that ellipticals formed
on timescales $\la 1 {\rm\,Gyr}$ (e.g.~Bender 1997).

Models for chemical enrichment during the formation of elliptical
galaxies support this conclusion but illustrate its sensitivity to
assumptions about stellar initial mass functions and supernovae yields
(Thomas, Greggio, \& Bender 1999).  With a Salpeter IMF, modest levels
of $\alpha$-process enrichment ($[{\alpha}/{\rm Fe}] \simeq 0.2$) can
result from hierarchical collapse with a $1 {\rm\,Gyr}$ star formation
timescale; within the uncertainties, mergers of newly-formed ($\la 3
{\rm\,Gyr}$ old) disk galaxies also work, while mergers of
present-epoch disks are excluded.  But drastic measures seem needed to
explain the enrichments of $[{\alpha}/{\rm Fe}] \simeq 0.5$ found in
nuclear disks of cluster ellipticals (Surma \& Bender 1995, Mehlert et
al.~1998).  Such high ratios suggest rapid enrichment in starbursts
dominated by massive stars; ``top-heavy'' IMFs are also indicated by
optical and IR photometry of ongoing starbursts (e.g.~Kennicutt 1998b,
p.~71).  These arguments must be weighed against the general evidence
for a universal IMF (e.g.~Elmegreen, these proceedings).  Studies of
$[{\alpha}/{\rm Fe}]$ in post-starburst merger remnants may clarify
the issue; significant $\alpha$-process enrichment would support the
case for a top-heavy IMF, while little or no enrichment would support
the case against recent disk galaxy mergers.

The hot gas in galaxy clusters, which apparently contains most of the
metals in these systems, may further constrain chemical enrichment
models for cluster ellipticals.  First, the sheer quantity of metals
in the intra-cluster medium seems to require a larger fraction of
stars becoming supernovae then predicted by a standard IMF (Renzini et
al.~1993).  Second, the ICM has nearly-solar $[{\alpha}/{\rm Fe}]$
ratios consistent with enrichment by both SN~II and SN~Ia (Ishimaru \&
Arimoto 1997).  If $\alpha$-process elements were somehow segregated
in galaxies, the ICM might be expected to show an {\it excess\/} of
Fe, which is not observed (cf. Mushotzky et al.~1996).

\subsection{Globular Clusters}

Young star clusters are observed in star-forming galaxies like the LMC
(Elson \& Fall 1985) and in intense starburst galaxies (Meurer et
al.~1995, Whitmore \& Schweizer 1995).  These clusters have half-light
radii of less than $5 {\rm\,pc}$, masses of $10^4$ to~$10^7
{\rm\,M_\odot}$, and metal abundances comparable to their parent
starbursts.  Their luminosity functions follow power laws with slopes
of $-1.6$ to~$-2$, intriguingly close to the mass function of giant
molecular clouds (Harris \& Pudritz 1994).  However, it's not entirely
clear that cluster luminosity is a good indicator of mass since some
range of cluster ages is usually present.

Cluster population correlates with galactic environment as well as
morphology (e.g.~Harris 1994).  In terms of the specific frequency
$S_{\rm N}$, defined as the number of globular clusters divided by the
galaxy luminosity in units of $M_{\rm V} = -15$, ellipticals in rich
cluster environments have $S_{\rm N} \simeq 4$ to $10$, field
ellipticals have $S_{\rm N} \simeq 2$, and spirals have $S_{\rm N}
\simeq 0.5$ to $1$.  Evidence is accumulating that globular cluster
populations can be augmented by merger-induced starbursts:

\begin{itemize}

\item
Ongoing and recent mergers (e.g.~NGC~4038/9, NGC~7252, NGC~3921) have
populations of blue luminous clusters with ages of less than $1
{\rm\,Gyr}$ (Whitmore \& Schweizer 1995, Schweizer et al.~1996, Miller
et al.~1997).

\item
Older remnants (e.g.~NGC~3610) have redder and fainter clusters with
ages of a few Gyr (Whitmore et al.~1997).

\item
Predicted specific frequencies in merger remnants increase to $S_{\rm
N} \simeq 2$ or~$3$ over $\sim 10 {\rm\,Gyr}$ as the stellar
populations fade (Schweizer et al.~1996, Miller et al.~1997).

\item
Globular clusters in many elliptical galaxies have bimodal or
multimodal color (metalicity) distributions (e.g.~Harris 1994).

\end{itemize}

These findings imply that the metal-rich globular clusters in field
ellipticals can form during mergers of disk galaxies and subsequently
assimilate into existing cluster populations (Ashman \& Zepf 1992).
This process is illustrated in Fig.~1, which plots specific
frequencies of metal rich ($S_{\rm N}^{\rm R}$) and metal-poor
($S_{\rm N}^{\rm P}$) globulars.  But the metal-{\it poor\/} globulars
in cluster ellipticals can't be explained in the same way (Forbes,
Brodie, \& Grillmair 1997).  Merging of metal-rich systems produces
metal-rich clusters; if cluster ellipticals owed their larger globular
populations to starbursts in metal-rich material, they would have
$S_{\rm N}^{\rm R} \gg S_{\rm N}^{\rm P}$, which is not observed.
This discrepancy is strongest for cD galaxies, which have $S_{\rm
N}^{\rm R} \ll S_{\rm N}^{\rm P} \simeq 10$.

\begin{figure}[t!]
\begin{center}
\plotfiddle{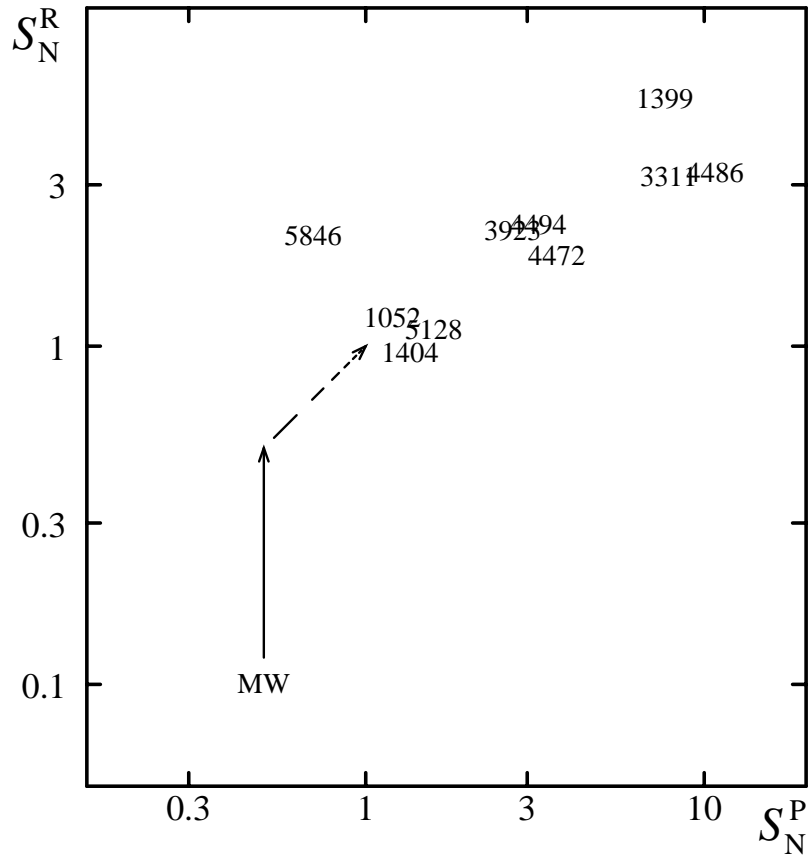}{10.5cm}{0}{90}{90}{-274}{-180}
\end{center}
\caption{Specific frequencies of metal-rich and metal-poor globular
clusters in a sample of elliptical galaxies (data from Forbes et
al.~1997).  The Milky Way (MW) is plotted assuming a ratio of
metal-rich to metal-poor clusters $N^{\rm R}/N^{\rm P} \simeq 0.2$.
The arrows show the result of a merger between two disk galaxies;
first the induced starburst forms metal-rich globulars, and then the
remnant fades, eventually attaining specific frequencies
characteristic of field ellipticals.}
\end{figure}

The question of high $S_{\rm N}$ in cluster ellipticals boils down to
this: fewer stars, or more globulars?  One way to get fewer stars is
to merge galaxies {\it after\/} their metal-poor globular clusters
have formed but before they build up substantial disks.  For example,
the Milky Way as it was $\sim 10 {\rm\,Gyr}$ ago could serve as a
building-block for cluster ellipticals; the halo of our galaxy,
considered alone, has $S_{\rm N}^{\rm P} \simeq 4$.  However, mergers
of Milky Way halos or dwarf elliptical galaxies (Miller et al.~1998)
still fall short of the high $S_{\rm N}^{\rm P}$ values of cD galaxies
and don't explain the high metalicities of luminous galaxies.  Another
way to end up with fewer stars is to eject most of the gas after the
initial epoch of globular cluster formation; the problem here is that
the ejection efficiency must be {\it higher\/} in cD galaxies, which
have the deepest potential wells and might be expected to retain the
most gas (Harris, Harris, \& McLaughlin 1998).

Alternately, the production of globulars may have been more efficient
in high-redshift starbursts.  Even at low-$z$, about $20$\% of the UV
emitted by starbursts comes from compact star-forming knots (Meurer et
al.~1995); if all these knots survived as clusters, the specific
frequency for a pure starburst population would be $S_{\rm N} \simeq
60$.  Moreover, these knots are concentrated where the surface
densities are highest.  It's likely that net yields of star clusters
increase rapidly with increasing gas pressure (Elmegreen \& Efremov
1997); ``highly crunched gas'' (Schweizer 1987) in early starbursts
might naturally produce even higher specific frequencies.

If so, then globular cluster systems reflect the starburst histories
of their parent galaxies: Large populations of metal-poor globulars
are due to efficient cluster production in early, low-metalicity
starbursts, while predominantly metal-rich systems (e.g.~NGC~5846)
formed more recently.  Metallicity distributions for globular cluster
populations support this idea; giant elliptical galaxies have a range
of distributions, often showing multiple peaks between $[{\rm Fe/H}]
\simeq -1.2$ and $0.2$ (Harris 1994).  This has been interpreted as
contradicting merger models and favoring a collapse in ``two distinct
phases'' with bursts of cluster formation separated by several Gyr
(Forbes et al.~1997).  But the observed variety of metalicity
distributions can arise naturally in a sequence of dissipative mergers
if each merger mixes up the available gas and contributes clusters of
fairly limited metalicity range to the cumulative population.  The
final distribution will then be a stochastic sum of a modest number of
peaks, in general accord with the observations.  Moreover, delays of
several Gyr between major starbursts are easily explained by merging,
but quite hard to explain as a result of internal events within a
single galaxy.  This picture for the origin of cluster populations
seems ripe for study with the semi-analytic tools for galaxy formation
in CDM (Kauffman, White, \& Guiderdoni 1993, Cole et al. 1994).

\section{Conclusions}

A wide range of circumstantial evidence suggests that merging played
an important role in galactic evolution long before the present epoch.
The key points of the argument can be summed up as follows:

\begin{itemize}

\item
Starbursts and AGN are signposts of high-redshift mergers; the high
incidence of such objects at $z \simeq 2$ to~$4$ reflects frequent
merging of juvenile galaxies.

\item
The bulk of the Milky Way's halo merged more than $10 {\rm\,Gyr}$ ago
as part of this activity.

\item
Cluster ellipticals were formed in dissipative mergers before $z
\simeq 2$; their immediate progenitors were few and only moderately
gassy.  Early-type galaxies in the field and in small groups have
continued forming right up to the present, most recently via
dissipative merging of disk galaxies.

\item
Globular cluster populations in cluster and field ellipticals differ
systematically because the former merged first.  The metal-rich
globular clusters of all ellipticals are relics of their final
dissipative mergers.

\end{itemize}

Peering back to redshifts $z \sim 5$ and greater, the familiar Hubble
types vanish altogether, and the universe is populated with numerous
``subgalaxies'' apparently much smaller than present-epoch galaxies.
Merging, first between such subgalaxies, and subsequently between more
familiar objects, played a key role in assembling the galaxies we know
today.  Despite the seductiveness of biological metaphors, it may be
misleading to say that early-type galaxies are the survivors of an
evolutionary process.  Rather, early-type galaxies emerged from a
merger-dominated environment; in some sense, they are {\it fixed
points\/} of merging transformations.

\acknowledgments

I thank Alex Stephens and Hector Vel\'azquez for communicating results
in advance of publication, and John Barnes for communicating Wilford's
article shortly after publication.  I'm also grateful for discussions
with Nobuo Arimoto, Carlos Frenk, and David Spergel.  I thank Jun
Makino and the University of Tokyo for hospitality while I researched
this article.  My research made use of NASA's Astrophysics Data System
Abstract Service.


\begin{references}

\reference Abraham, R.G. 1999, in Galaxy Interactions at Low and
High Redshifts, eds J.E. Barnes \& D.B. Sanders (Kluwer, Dordrecht),
p. 11

\reference Ashman, K.M. \& Zepf, S.E. 1992, \apj, 384, 50

\reference Bahcall, J.N., Kirhakos, S., \& Schneider, D.P. 1995,
\apj, 447, L1

\reference Barger, A.J., Cowie, L.L., Sanders, D.B., Fulton,
E., Taniguchi, Y., Sato, Y., Kawara, K., \& Okuda, H. 1998, Nature,
394, 248

\reference Barger, A.J., Cowie, L.L., Trentham, N., Fulton, E., Hu,
E.M., Songaila, A., \& Hall, D. 1999, \aj, 117, 102

\reference Barnes, J.E. 1988, \apj, 331, 699

\reference Barnes, J.E. \& Hernquist, L. 1996, \apj, 471, 115

\reference Baugh, C.M., Cole, S., \& Frenk, C.S. 1996, \mnras, 283,
1361

\reference Begelman, M.C., Blandford, R.D., \& Rees, M.J. 1984,
Reviews of Modern Physics, 56, 255

\reference Bender, R. 1997, in The Nature of Elliptical Galaxies,
eds. M. Arnaboldi, G.S. Da Costa, \& P. Saha (ASP, San Francisco),
p. 11

\reference Bender, R. \& Surma, P. 1992, AA, 258, 250

\reference Bernardi, M., Renzini, A., da Costa, L.N., Wegner, G.,
Alonso, M.V., Pellegrini, P.S., Rit\'e, C., \& Willmer, C.N.A. 1988,
\apj, 508, L43

\reference Blitz, L. Spergel, S.N., Teuben, P.J., Hartmann, D., \&
Burton, W.B. 1999, \apj, 514, 000

\reference Boyce, P.J., Disney, M.J., Blades, J.C.,
Boksenberg, A., Crane, P., Deharveng, J.M., Macchetto, F.D., Mackay,
C.D., \& Sparks, W.B. 1996, \apj, 473, 760

\reference Brinchmann, J., Abraham, R., Schade, D.,
Tresse, L., Ellis, R. S., Lilly, S., Le Fevre, O., Glazebrook, K.,
Hammer, F., Colless, M., Crampton, D., \& Broadhurst, T. 1998, \apj,
499, 112

\reference Blumenthal, G.R., Faber, S.M., Primack, J.R., Rees,
M.J. 1984, Nature, 311, 517

\reference Carollo, M., Franx, M., Illingworth, G.D., \& Forbes,
D.A. 1997, \apj, 481, 710

\reference Chambers, K.C., Miley, G.K., \& van Breugel,
W.J.M. 1987, Nature, 329, 604

\reference Cole, S., Aragon-Salamanca, A., Frenk, C.S., Navarro, J.F.,
\& Zepf, S.E. 1994, \mnras, 271, 781

\reference Davies, R.L., Sadler, E.M., Peletier, R.F. 1993, \mnras,
262, 650

\reference de Carvalho, R.R. \& Djorgovski, S. 1992, \apj, 389,
L49

\reference Disney, M.J., Boyce, P.J., Blades, J.C.,
Boksenberg, A., Cane, P., Deharveng, J.M., Macchetto, F., Mackay,
C.D., Sparks, W.B., \& Phillipps, S. 1995, Nature, 376, 150

\reference Eggen, O.J. 1987, in The Galaxy, eds. G. Gilmore \&
B. Carswell (Reidel, Dordrecht), p. 211

\reference Elmegreen, B.G. \& Efremov, Y.N. 1997, \apj, 480, 235

\reference Elson, R.A. \& Fall, S.M. 1985, \pasp, 97, 692

\reference Faber, S.M., Trager, S.C., Gonzalez, J.J., \&
Worthey, G. 1994, in Stellar Populations, eds. P.C. van der Kruit \&
G. Gilmore (Kluwer, Dordrecht), p. 249

\reference Fisher, K.B., Bahcall, J.N., Kirhakos, S., \&
Schneider, D.P. 1996, \apj, 468, 469

\reference Forbes, D.A., Brodie, J.P., \& Grillmair, C.J. 1997,
AJ, 113, 1652

\reference Franx, M. \& Illingworth, G.D. 1988, \apj, 327, L55

\reference Freeman, K.C. 1993, in Galaxy Evolution: The Milky Way
Perspective, ed. S.R. Majewski (ASP, San Francisco), p. 125

\reference Gilmore, G. \& Wyse, R.F.G. 1998, \aj, 116, 748

\reference Gilmore, G., Wyse, R.F.G., \& Kuijken, K. 1989, ARAA, 27,
555

\reference Harris, W.E. 1994, in Stellar Populations, eds. P.C. van
der Kruit \& G. Gilmore (Kluwer, Dordrecht), p. 85

\reference Harris, W.E., Harris, G.L.H., \& McLaughlin,
D.E. 1998, \aj, 115, 1801

\reference Harris, W.E. \& Pudritz, R.E. 1994, \apj, 429, 177

\reference Heckman, T.M. 1998, astro-ph/9801155

\reference Heckman, T.M., Smith, E.P., Baum, S.A., van
Breugel, W.J., Miley, G.K., Illingworth, G.D., Bothun, G.D., \&
Balick, B. 1986, \apj, 311, 526

\reference Hernquist, L. \& Barnes, J.E. 1991, Nature, 354, 210

\reference Hibbard, J.E. \& Mihos, J.C. 1995, \aj, 110, 140

\reference Hibbard, J.E. \& Vacca, W.D. 1997, \aj, 114, 1741

\reference Huang, S. \& Carlberg, R.G. 1997, \apj, 480, 503

\reference Hughes, D.H., Serjeant, S., Dunlop, J.,
Rowan-Robinson, M., Blain, A., Mann, R.G., Ivison, R., Peacock, J.,
Efstathiou, A., Gear, W., Oliver, S.,\\
Lawrence, A., Longair, M., Goldschmidt, P., \& Jenness, T. 1998,
Nature, 394, 241

\reference Ibata, R.A., Gillmore, G., \& Irwin, M.J. 1995, \mnras,
277, 781

\reference Ishimaru, Y. \& Arimoto, N. 1997, PASJ, 49, 1

\reference Kauffmann, G. 1996, \mnras, 281, 487

\reference Kauffmann, G., Charlot, S., \& White, S.D.M. 1996, \mnras,
283, L117

\reference Kauffmann, G., White, S.D.M., \& Guiderdoni, B. 1993,
\mnras, 264, 201

\reference Kennicutt, R. 1998a, \apj, 498, 541

\reference Kennicutt, R. 1998b, in Galaxies: Interactions and Induced
Star Formation, eds. D. Friedli, L. Martinet, \& D. Pfenniger
(Springer, Berlin), p. 1

\reference Knopp, G.P. \& Chambers, K.C. 1997, \apj, 487, 644

\reference Kormendy, J. 1990, in The Edwin Hubble Centennial
Symposium: The Evolution of the Universe of Galaxies, ed. R. G. Kron
(ASP, San Francisco), p. 33

\reference Kormendy, J. \& Sanders, D.B. 1992, \apj, 390, L53

\reference Lacey, C. \& Cole, S. 1993, \mnras, 262, 627

\reference Layzer, D. 1954, \aj, 59, 170

\reference Lynden-Bell, D. \& Lynden-Bell, R.M. 1995, \mnras,
275, 429

\reference Majewski, S.R. 1996, \apj, 459, L73

\reference Majewski, S.R., Munn, J.A., \& Hawley, S.L. \apj, 427, L37

\reference Makino, J. \& Hut, P. 1997, \apj, 481, 83

\reference Mathewson, D.S. 1985, Proc. Astron. Soc. Australia, 6, 104

\reference McCarthy, P.J., van Breugel, W.J.M., Spinrad, H., \&
Djorgovski, S.G.  1987, \apj, 321, L29

\reference Mehlert, D., Saglia, R.P., Bender, R., \& Wegner,
G. 1998, AA, 332, 33

\reference Meurer, G.R., Heckman, T.M., Leitherer, C., Kinney,
A., Robert, C., \& Garnett, D.R. 1995, \aj, 110, 2665

\reference Miller, B.W., Lotz, J.M., Ferguson, H.C., Stiavelli,
M., \& Whitmore, B.C. 1998, \apj, 508, L133

\reference Miller, B.W., Whitmore, B.C., Schweizer, F., \& Fall,
S.M. 1997, \aj, 114, 2381

\reference Mo, H.J., Mao, S., \& White, S.D.M. 1998, \mnras, 295,
319

\reference Mushotzky, R., Loewenstein, M., Arnaud, K.A.,
Tamura, T., Fukazawa, Y., Matsushita, K., Kikuchi, K., \& Hatsukade,
I. 1996, \apj, 466, 686

\reference Pettini, M., Kellog, M., Steidel, C.C., Dickinson,
M., Adelberger, K.L., \& Giavalisco, M. 1998a, \apj, 508, 539

\reference Pettini, M., Steidel, C.C., Dickenson, M., Kellog, M.,
Giavalisco, M., \& Adelberger, K.L. 1998b, in AIP Conf. Proc. 408,
Ultraviolet Universe at Low and High Redshift, ed. W. Waller (AIP, New
York), p. 279

\reference Preston, G.W., Beers, T.C., \& Schectman, S.A. 1994,
AJ, 108, 538

\reference Rush, B., McCarthy, P.J., Athreya, R.M., \& Persson,
S.E. 1997, \apj, 484, 163

\reference Sanders, D.B. \& Mirabel, I.F. 1996, ARAA, 34, 749

\reference Sarajedini, A., Chaboyer, B., \& Demarque, P. 1997, \pasp,
109, 1321

\reference Schwarz, M.P. 1981, \apj, 247, 77

\reference Schweizer, F. 1982, \apj, 252, 455

\reference Schweizer, F. 1987, in Nearly Normal Galaxies,
ed. S.M. Faber (Springer-Verlag, Berlin), p. 18

\reference Schweizer, F. 1990, in Dynamics and Interactions of
Galaxies, ed. R. Wielen (Springer, Berlin), p. 60

\reference Schweizer, F. 1998, in Galaxies: Interactions and
Induced Star Formation, eds. D. Friedli, L. Martinet, \& D. Pfenniger
(Springer, Berlin), p. 105

\reference Schweizer, F., Miller, B.W., Whitmore, B.C., \& Fall,
S.M. 1996, \aj, 112, 1839

\reference Schweizer, F. \& Seitzer, P. 1992, \aj, 104, 1039

\reference Schweizer, F., Seitzer, P., Faber, S.M.,
Burstein, D., Dalle Ore, C.M., \& Gonzalez, J.J. 1990, \apj, 364, L33

\reference Searle, L. \& Zinn, R. 1978, \apj, 225, 357

\reference Somerville, R.S., Primack, J.R., Faber, S.M. 1998,
astro-ph/9806228

\reference Steidel, C.C., Giavalisco, M., Dickinson, M., \&
Adelberger, K.L. 1996a, \aj, 112, 352

\reference Steidel, C.C., Giavalisco, M., Pettini, M.,
Dickinson, M., \& Adelberger, K.L. 1996b, \apj, 462, L17

\reference Stephens, A. 1999, \aj, 117, 000

\reference Stetson, P.B., VandenBerg, D.A., \& Bolte, M. 1996, \pasp,
108, 560

\reference Stockton, A. 1999, in Galaxy Interactions at Low and High
Redshift, eds. J.E. Barnes \& D.B. Sanders (Kluwer, Dordrecht), p. 311

\reference Stockton, A., Canalizo, G., \& Close, L.M. 1998, \apj,
500, L121

\reference Stockton, A. \& Mackenty, J.W., 1983, Nature, 305, 678

\reference Stockton, A. \& Ridgway, S.E. 1991, \aj, 102, 488

\reference Stockton, A. \& Ridgway, S.E. 1998, \apj, 115, 1340

\reference Surma, P. \& Bender, R. 1995, AA, 298, 405

\reference Renzini, A., Ciotti, L., D'Ercole, A., \&
Pellegrini, S. 1993, \apj, 419, 52

\reference Rix, H.-W. \& White, S.D.M. 1992, \mnras, 254, 389

\reference Tadhunter, C.N., Fosbury, K.R.A.E., \& di Serego
Alighieri, S. 1989, in BL Lac Objects, eds. L. Maraschi, T. Maccaro,
\& M.H. Ulrich (Springer, Berlin), p. 79

\reference Thomas, D., Greggio, L., \& Bender, R. 1999, \mnras, 302,
537

\reference Toomre, A. \& Toomre, J. 1972, \apj, 178, 623

\reference Toomre, A. 1977, in The Evolution of Galaxies and of
Stellar Populations, eds. B.M. Tinsley \& R.B. Larson (Yale
Observatory, New Haven), p. 401

\reference T\'oth, G. \& Ostriker, J.P. 1992, \apj, 389, 5

\reference Unavane, M., Wyse, R.F.G., \& Gilmore, G. 1996, \mnras,
278, 727

\reference van den Bergh, S. 1993, \aj, 105, 971

\reference van den Bergh, S., Abraham, R.G., Ellis, R.S.,
Tanvir, N.R., Santiago, B.X., \& Glazebrook, K.G. 1996, \aj, 112, 359

\reference van Dokkum, P.G., Franx, M., Kelson, D.D., \&
Illingworth, G.D. 1998, 504, L17

\reference Vel\'azquez, H. \& White, S.D.M. 1998, \mnras, in press

\reference Walker, I.R., Mihos, J.C., \& Hernquist, L. 1996, \apj,
460, 121

\reference White, S.D.M. \& Rees, M.J. 1978, \mnras, 183, 341

\reference Weedman, D.W., Wolovitz, J.B., Bershady, M.A., \&
Schneider, D.P. 1998, \aj, 116, 1643

\reference Whitmore, B.C., Miller, B.W., Schweizer, F., \& Fall,
S.M. 1997, \aj, 114, 1797

\reference Whitmore, B.C. \& Schweizer, F. 1995, \aj, 109, 960

\reference Wilford, J.N. 1998, The New York Times, November 24, p. F5

\reference Wilson, A.S. \& Colbert E.J.M. 1995, \apj, 438, 62

\reference Worthey, G., Faber, S.M., \& Gonzalez, J.J. 1992, \apj,
398, 69

\reference Zepf, S.E. \& Koo, D.C. 1989, \apj, 337, 34

\end{references}
\end{document}